# Non-radiative exciton energy transfer in hybrid organic-inorganic heterostructures

S. Chanyawadee[1], P. G. Lagoudakis[1*], R. T. Harley[1], D. G. Lidzey[2] and M. Henini[3]

[1] School of Physics & Astronomy, University of Southampton, Southampton, SO17 1BJ, UK

[2] Department of Physics and Astronomy, The University of Sheffield, Sheffield, S3 7RH, UK

[3] School of Physics and Astronomy, University of Nottingham, Nottingham, NG7 2RD, UK

Non-radiative optical energy transfer from a GaAs quantum well to a thin overlayer of an infrared organic semiconductor dye is unambiguously demonstrated. The dynamics of exciton transfer are studied in the time-domain using pump-probe spectroscopy at the donor site and fluorescence spectroscopy at the acceptor site. The effect is observed as simultaneous increase of the population decay rate at the donor and of the rise time of optical emission at the acceptor sites. The hybrid configuration under investigation provides an alternative non-radiative, non-contact pumping route to electrical carrier injection that overcomes the losses imposed by the associated low carrier mobility of organic emitters.

---

[*] Corresponding author. Email: pavlos.lagoudakis@soton.ac.uk

The brightness and flexibility of organic semiconductors have rendered them promising new materials for the optoelectronics industry [1, 2] with full colour displays and flexible screens being a few of the upcoming applications. Although organic emitters are in general brighter than their inorganic counterparts, the unresolved issue of low carrier mobility in organic compounds poses limitations on the total efficiency of these devices [3, 4] and is partly responsible for the unattainable of an electrically pumped organic laser. A route to circumvent altogether issues associated to low carrier mobility in organic emitters is to engineer devices that utilise alternative pumping schemes to electrical injection and transport while still benefiting from the brightness of the organic dyes. By tailoring the optoelectronic properties of a hybrid organic-inorganic material system, carriers could be injected and transported into the inorganic part and be subsequently transferred in a non-contact way directly to the organic emitter. Rapid and efficient energy transfer can occur either by coherently coupling the inorganic and inorganic semiconductors in the strong coupling regime [5-7] or by non-radiative energy transfer from an inorganic semiconductor slab to an organic overlayer [8, 9]. In the latter process, no real photons are involved and the efficiency of the transfer process between donor and acceptor sites, first studied by Förster [10], can exceed that of radiative energy transfer routinely used in phosphor light emitting devices [11]. First experimental evidence of the above mechanism was observed in hybrid structures between Wannier excitons in wide band gap semiconductor quantum wells (donors) and colloidal semiconductor quantum dots [12, 13] (acceptors) or Frenkel excitons in organic fluorophores [14-16].

Time-resolved studies in hybrid quantum well heterostructures have demonstrated the effect of an extra decay channel in the quantum well emission (donor) [12-14]. However, this alone cannot be taken as evidence of energy transfer to the acceptor sites because the deposition of an overlayer of organic dye molecules, fluorophores or colloidal quantum dots on the quantum well barrier separating donor and acceptor sites could lead to modification of its surface states, which would itself modify the decay dynamics. Any changes in the photoluminescence decay of a quantum well in a hybrid structure can only be conclusively attributed to non-radiative energy transfer to the acceptor site if the effect of energy transfer is also observed directly in the emission properties of the acceptor.

In this letter, we demonstrate non-radiative energy transfer between a single GaAs quantum well donor and an acceptor of an organic dye both at the donor's and acceptor's sites. The dynamics of exciton transfer are studied in the time-domain using pump-probe reflection spectroscopy at the donor site and time-resolved fluorescence decay measurements at the acceptor site. Furthermore, we investigate the dependence of the energy transfer efficiency on the spectral overlap between donor and acceptor by varying the width of the single quantum well donor. The precise growth of the GaAs heterostructures allows us to controllably modify the spectral overlap between donor and acceptor in the hybrid structure while accurately controlling the distance between quantum well excitons and organic molecules.

The organic/inorganic hybrid structures consist of a GaAs/AlGaAs single quantum well (donor) and an organic dye overlayer (acceptor) as shown in Fig. 1(a). The organic material is dissolved in methanol and spin coated on the quantum well to form a uniform film with a nominal thickness of 12 nm [Fig. 1(b)]. By varying the concentration of the initial solution of the dye we control the thickness of the organic overlayer with ±15 nm accuracy [inset in Fig. 1(b)]. The infra-red emitting dye that we have used [17] when dissolved in methanol and spin coated on a glass substrate has an absorption spectrum as shown in Fig. 1(c) (dashed-dotted curve, red online). The single GaAs/Al$_{0.35}$Ga$_{0.65}$As quantum well structures are grown by molecular beam epitaxy on (100) GaAs substrates with a 500 nm GaAs buffer layer. In sample A the quantum well width is 4 nm, while in sample B the width is 6.5 nm. For both structures the width of the inner barrier is 20 nm, with the top barrier (which also controls the separation between quantum well excitons and organic molecules), being 7.5 nm. The photoluminescence spectrum of the two single quantum well structures at 25 K is shown in Fig. 1(c). Sample A has a photoluminescence peak at 772 nm (dashed curve) while sample B has a photoluminescence peak at 795 nm (solid curve, red online). Thus by varying the thickness of the single quantum well the spectral overlap between the photoluminescence of the quantum well and the absorption of the organic overlayer can be controllably tuned.

The broad fluorescence spectrum of the organic emitter and its overlap with the photoluminescence spectrum of the quantum well makes it impossible to study both donor and acceptor using photoluminescence alone. Therefore, we use pump-probe reflectivity measurements to study the transfer dynamics from the quantum wells to the organic overlayer combined with spectrally resolved time-correlated single photon counting measurements of the fluorescence from the organic dye. Pump-probe reflectivity measurements have allowed for characterisation of the transient dynamics of excitons in the quantum wells while being insensitive to the presence of a thin organic overlayer. A schematic of the pump-probe set-up is shown in Fig. 2(a). A tunable mode-locked Ti:sapphire oscillator with 76 MHz repetition rate and ~2 ps pulse width is used to excite the quantum wells resonant with the n=1 heavy hole excitonic absorption. The pump-induced change of the reflected probe intensity ($\Delta R$) is measured by scanning the time delay between pump and probe pulses. Small pump-induced changes of the reflected probe beam are measured using balanced photodiodes and a lock-in detection scheme [18].

Figure 2(b), (c) show the population decay ($\Delta R$) of the quantum wells only (solid squares) and the population decay of quantum wells in the presence of the organic overlayer (open circles, red online). The excitation wavelengths are 768 nm and 792 nm for sample A and B respectively. The faster depopulation in the hybrid organic-inorganic structure demonstrates that the presence of the organic layer results in an additional decay channel. The decay time for sample A decreases from 326 ps to 235 ps whereas that for sample B decreases from 642 ps to 389 ps and the additional decay rate in the hybrid configuration, $k_{ET}$, can be derived from the photoluminescence decay rate of the hybrid structure ($k_H = \tau_H^{-1}$), and of the bare QW ($k_{QW} = \tau_{QW}^{-1}$) from $k_{ET} = k_H - k_{QW}$.

Alone, this result does not demonstrate that the onset of the extra decay channel in the quantum well emission is due to energy transfer to the organic overlayer. To do this, we directly compare the time-evolution of the fluorescence of the organic dyes in the hybrid structures and on glass substrates where no energy transfer occurs, using spectrally filtered, time-correlated single photon counting. Figure 3 compares the fluorescence on a glass substrate (solid squares) with that on sample A (solid circles, red online). Here, both structures are excited at 745 nm and spectral filters are used to select the fluorescence of the organic dyes only. Evidently, the emission rise time from the hybrid structure is considerably slower, which clearly indicates energy transfer from the quantum well excitons to the organic dye.

Following the direct observation of the transfer to the organic overlayer, the extra rate, $k_{ET}$, estimated from the pump-probe measurements on the quantum wells can be confidently assigned to non-radiative energy transfer. The efficiency of the process is derived from $\eta = k_{ET}/(k_{ET} + k_{QW})$, which for the samples A and B is $\eta_A = 0.23$ and $\eta_B = 0.39$ respectively. Theoretically the energy transfer rate scales linearly with the spectral overlap between donor and acceptor [19]. We therefore expect here the ratio of the transfer rate of the two hybrid structures, $k_A/k_B = 1.18 \pm 0.21$, to be equal to the ratio of the corresponding spectral overlaps of the two quantum wells (samples A and B) with the organic dye absorption spectrum. From the data of Fig. 1(c) we obtain for this ratio $\Omega_A/\Omega_B = 1.27$ which is indeed reasonably close to the ratio of transfer rates. This agreement further confirms the nature of the process and the validity of the pump-probe technique to characterise non-radiative energy transfer in hybrid structures.

To quantitatively correlate the results of non-radiative energy transfer at the donor and acceptor sites we use a simple kinetic model to describe exciton dynamics in the quantum well and the organic layer:

$$\frac{dN_{QW}}{dt} = -k_{QW}N_{QW} - k_{ET}N_{QW} \tag{1}$$

$$\frac{dN_{OG1}}{dt} = -k_{OG1}N_{OG1} + k_{ET}N_{QW} \tag{2}$$

$$\frac{dN_{OG2}}{dt} = -k_{OG2}N_{OG2} + k_{ET}N_{QW} \tag{3}$$

where $N_{QW}$ and $N_{OG1,2}$ are exciton populations in the quantum well and the organic overlayer respectively, and $k_{OG1}=1\times10^{10}$ s$^{-1}$, $k_{OG2}=8.3\times10^{8}$ s$^{-1}$ are the fluorescence decay rates of the bare organic dye as derived from the fast and slow components of its almost biexponential decay (Fig. 3, solid squares). The analytic solution of the above set of equations for the fluorescence decay of the organic dye is

$$I(t) = \sum_{x=1,2}\frac{C_x N_{QW}(t=0)k_{ET}}{k_{OGx} - k_{QW} - k_{ET}}(e^{-(k_{QW}+k_{ET})t} - e^{-k_{QW}t}) \tag{4}$$

where $x=1,2$ and $C_{1,2}$ are constants. The dashed curve on Fig. 3 is a fit of the measured fluorescence decay (solid circles, red online) in the hybrid configuration using Eq. (4) (with $C_1/C_2 = 4$ expressing the contribution of the fast and slow component of the organic dye on the fluorescence decay). Considering that the photoluminescence decay rate, $k_{QW}$, the non-radiative energy transfer rate, $k_{ET}$, and the fluorescence decay rates, $k_{OG1}$, $k_{OG2}$, are not free parameters but are taken from the pump-probe measurements and the fluorescence decay of the organic dye on glass respectively, there is a very good agreement between this simple model and the experimental observation.

In the rapidly developing field of hybrid optoelectronics, where organic and inorganic semiconductors are merged in novel configurations that exploit the advantages of the individual compounds, non-radiative energy transfer is shown to play an important role. It allows for funnelling of energy in a non-contact, non-radiative way that overcomes limitation imposed by the low carrier mobility in organic semiconductors. The measurements we have described provide conclusive evidence of non-radiative energy transfer from a semiconductor quantum well to an overlayer of organic dye in the near-infrared regime by resolving the processes both at the donor and acceptor site thereby excluding other non-radiative mechanisms that could influence the photoluminescence properties of the quantum well only. Furthermore, by appropriate engineering of the semiconductor materials we observe high non-radiative energy transfer efficiencies (23% sample A and 39% sample B) that indicate the applicability of the process in real-world hybrid optoelectronic devices. Future work will address the issue of electrically-generated emission from such devices, and will address the realization of a hybrid organic-inorganic injection laser-diode.

The authors acknowledge Julia Bricks and Jurii Slominskii from the Institute of Organic Chemistry in Ukraine for the provision of the infra-red emitting organic dye. This work was supported by the University of Southampton Annual Adventure in Research Grant A2005/18, the Royal Society Research Grant No. 2006/R2, the EPSRC grant EP/F013876/1 and the Royal Thai Government.

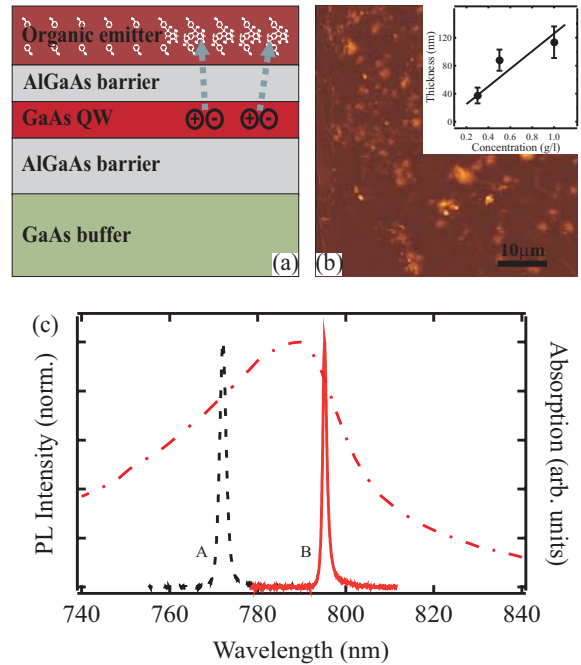

**FIG. 1** (color online). (a) Schematic diagram of organic/inorganic hybrid structures. Organic dye is deposited on 4.0 (sample A) and 6.5 nm thick quantum wells (sample B). (b) Atomic force microscopy image of organic dye layer deposited on a glass substrate. Inset: thickness of the organic film vs concentration of organic dye. (c) Spectral overlaps between the emission of sample A (dashed curve) and of sample B (solid curve, red online) at 25 K and the absorption of organic dye (dashed-dotted curve, red online).

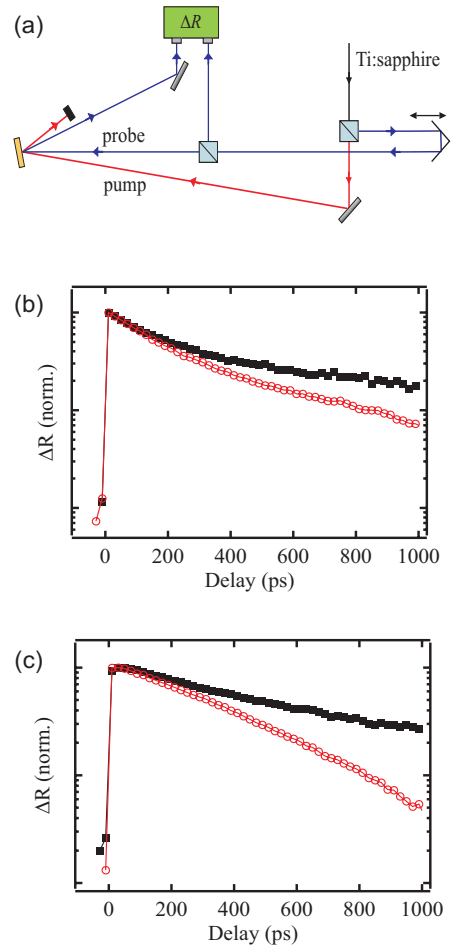

**FIG. 2** (color online). (a) A schematic of the pump-probe set-up. Normalized population decays of the bare quantum wells (solid squares) and the quantum wells in the hybrid structures (open circles, red online) for sample A (b) and sample B (c).

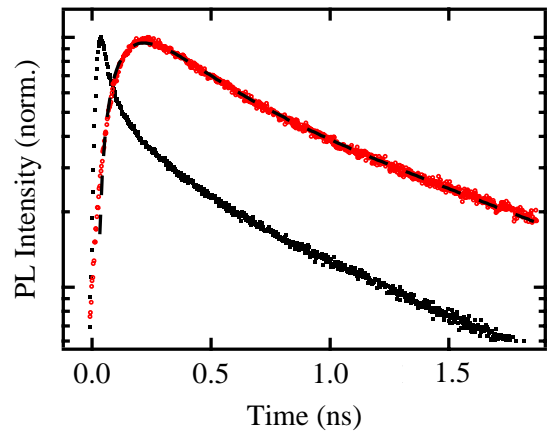

**FIG. 3** (color online). Photoluminescence decays of organic dye deposited on a glass substrate (solid squares) and on a single quantum well (sample A, open circles, red online). The dashed curve is the fit using Eq. (4).

[1]    H. Kanno *et al.*, Advanced Materials **18**, 339 (2006).

[2]    M. Segal *et al.*, Nature Materials **6**, 374 (2007).

[3]    P. S. Davids *et al.*, Applied Physics Letters **69**, 2270 (1996).

[4]    G. D. Sharma, V. S. Choudhary, and M. S. Roy, Solar Energy Materials and Solar Cells **91**, 1087 (2007).

[5]    J. Wenus *et al.*, Physical Review B **74** (2006).

[6]    V. Agranovich, H. Benisty, and C. Weisbuch, Solid State Communications **102**, 631 (1997).

[7]    R. J. Holmes *et al.*, Physical Review B **74** (2006).

[8]    D. Basko *et al.*, Eur. Phys. J. B **8**, 353 (1999).

[9]    V. M. Agranovich *et al.*, J. Phys. Condens. Matter **10**, 9369 (1998).

[10]    T. Förster, Annalen der Physik **2**, 55 (1948).

[11]    M. Achermann *et al.*, Nano Letters **6**, 1396 (2006).

[12]    M. Achermann, Nature **429**, 642 (2004).

[13]    S. Rohrmoser *et al.*, Applied Physics Letters **91** (2007).

[14]    G. Heliotis, Adv. Mater. **18**, 334 (2006).

[15]    S. Blumstengel *et al.*, Phys. Rev. Lett. **97**, 237401 (2006).

[16]    G. Itskos *et al.*, Physical Review B **76** (2007).

[17]    J. Wenus *et al.*, Organic Electronics **8**, 120 (2007).

[18]    R. T. Harley, O. Z. Karimov, and M. Henini, Journal of Physics D-Applied Physics **36**, 2198 (2003).

[19]    M. Achermann *et al.*, Journal of Physical Chemistry B **107**, 13782 (2003).